\def\ts     {\thinspace}
\def\kms    {\ifmmode{{\rm \ts km\ts s}^{-1}}\else{\ts km\ts s$^{-1}$}\fi}
\def\msol   {\ifmmode{{\rm M}_{\odot} }\else{M$_{\odot}$}\fi}
\def\lsol   {\ifmmode{L_{\odot}}\else{$L_{\odot}$}\fi}
\def\lfir   {\ifmmode{L_{\rm FIR}}\else{$L_{\rm FIR}$}\fi}
\def\zsol   {\ifmmode{{\rm Z}_{\odot}}\else{Z$_{\odot}$}\fi}

\def\aco    {\ifmmode{{\rm CO}(J\!=\!1\! \to \!0)}\else{{\rm CO}($J$=1$\to$0)}\fi}
\def\bco    {\ifmmode{{\rm CO}(J\!=\!2\! \to \!1)}\else{{\rm CO}($J$=2$\to$1)}\fi}
\def\cco    {\ifmmode{{\rm CO}(J\!=\!3\! \to \!2)}\else{{\rm CO}($J$=3$\to$2)}\fi}
\def\dco    {\ifmmode{{\rm CO}(J\!=\!4\! \to \!3)}\else{{\rm CO}($J$=4$\to$3)}\fi}
\def\eco    {\ifmmode{{\rm CO}(J\!=\!5\! \to \!4)}\else{{\rm CO}($J$=5$\to$4)}\fi}
\def\fco    {\ifmmode{{\rm CO}(J\!=\!6\! \to \!5)}\else{{\rm CO}($J$=6$\to$5)}\fi}
\def\gco    {\ifmmode{{\rm CO}(J\!=\!7\! \to \!6)}\else{{\rm CO}($J$=7$\to$6)}\fi}
\def\hco    {\ifmmode{{\rm CO}(J\!=\!8\! \to \!7)}\else{{\rm CO}($J$=8$\to$7)}\fi}
\def\ico    {\ifmmode{{\rm CO}(J\!=\!9\! \to \!8)}\else{{\rm CO}($J$=9$\to$8)}\fi}
\def\jco    {\ifmmode{{\rm CO}(J\!=\!10\! \to \!9)}\else{{\rm CO}($J$=10$\to$9)}\fi}
\def\kco    {\ifmmode{{\rm CO}(J\!=\!11\! \to \!10)}\else{{\rm CO}($J$=11$\to$10)}\fi}
\def\ci     {\ifmmode{{\rm C}{\rm \small I}}\else{C\ts {\scriptsize I}}\fi}
\def\hi     {\ifmmode{{\rm H}{\rm \small I}}\else{H\ts {\scriptsize I}}\fi}
\def\hh     {\ifmmode{{\rm H}_2}\else{H$_2$}\fi}
\def\cone {\ifmmode{{\rm C}{\rm \small I}(^3\!P_1\!\to^3\!P_0)}
     \else{C\ts {\scriptsize I}{\small$(^3\!P_1\!\to^3\!\!\!P_0)$}}\fi}
\def\ctwo {\ifmmode{{\rm C}{\rm \small I}(^3\!P_2\!\to^3\!P_1)}
     \else{C\ts {\scriptsize I}{\small$(^3\!P_2\!\to^3\!\!\!P_1)$}}\fi}
\def\cij {\ifmmode{{\rm C}{\rm \small I}\,(^3P_i\to^3P_j)}
\else{C\ts {\scriptsize I}\,{\small$(^3P_i\to^3P_j)$}}\fi}
\def\cii    {\ifmmode{{\rm C}{\rm \small II}}\else{C\ts {\scriptsize II}}\fi}
\def\tex {\ifmmode{{T}_{\rm ex}}\else{$T_{\rm ex}$}\fi}
\def\tmb {\ifmmode{{T}_{\rm mb}}\else{$T_{\rm mb}$}\fi}
\def\tkin {\ifmmode{{T}_{\rm kin}}\else{$T_{\rm kin}$}\fi}
\def\microns {\ifmmode{\mu{\rm m}}\else{$\mu$m}\fi}
\def\nhh   {\ifmmode{n({\rm H}_2)}\else{$n$(H$_2$)}\fi}
\def\gradv {\ifmmode{(dv/dr)}\else{$(dv/dr)$}\fi}
\def\CO10{{\hbox {CO(1--0)}}}

\def\,{\thinspace}
\def\Msun{M$_\odot$}

\def\Lsun{L$_\odot$}

\documentclass[twocolumn]{aa}    
\usepackage{graphicx}
\begin{document}
\titlerunning{Arp 220-West Black Hole}
\title{Black hole in the West Nucleus of Arp~220}
   \author{
          D. Downes
          \inst{1}
          \and
          A. Eckart
          \inst{2,3}
          }
   \institute{Institut de Radio Astronomie Millim\'etrique,
              Domaine Universitaire, 38406 St-Martin-d'H\`eres, France
         \and
             I.Physikalisches Institut, Universit\"at zu K\"oln, 
             Zulpicherstrasse 77,  50937 Cologne, Germany	 
         \and 
             Max-Planck-Institut f\"ur Radioastronomie, 
             Auf dem H\"ugel 69, 53121 Bonn, Germany
             }

   \date{Received 15 February 2007 / Accepted 13 March 2007}
   \abstract{
We present new observations with the IRAM Interferometer, 
in its longest-baseline configuration, of
the CO(2--1) line and the 1.3\,mm dust radiation 
from the Arp 220 nuclear region.  The dust source in the West nucleus 
has a size of 0.19$''\times 0.13''$ and 
 a 1.3\,mm brightness temperature of 90\,K.
This implies that the dust ring in the West nucleus 
has a high opacity, with $\tau = 1$ at 1.1\,mm.  
Not only is the dust ring itself optically thick in the 
submm and far-IR, but it is surrounded by the previously-known, 
rapidly rotating
molecular disk of size 0.5$''$ that is also optically thick in the 
mid-IR. The molecular ring is cooler than the hot dust disk because 
the CO(2--1) line is seen in absorption against the dust disk.
The dust ring is massive (10$^9$\,\Msun ), compact (radius 35\,pc), 
and hot (true dust temperature 170\,K). It resembles rather strikingly 
the dust ring detected around the quasar APM 08279+52, and is most 
unlike the warm, extended dust sources in starburst galaxies. Because there is 
a strong temperature gradient from the hot dust ring to the 
cooler molecular disk, the heating must come from a concentrated source, 
an AGN accretion disk that is   
completely invisible at optical wavelengths, and heavily obscured
in hard X-rays.

}
   \maketitle
\section{Evidence for a black hole, so far.} 
Evidence is growing for a supermassive black hole in the
West nucleus of the Arp 220 merger:

\noindent
1) {\it X-rays:} The {\it Chandra} point source in the
West nucleus has a 2--10 keV luminosity of
$10^7$\,\Lsun, or 10$^{-5}$ times the FIR luminosity.
Extended hard X-ray emission in the vicinity raises the total nuclear
X-ray luminosity by an order of magnitude, and one 
cannot rule out a much greater X-ray flux from an AGN hidden behind
H$_2$ column densities $>5\times 10^{24}$\,cm$^{-2}$
(Clements et al.\ 2002; Ptak et al.\ 2003).  Updated astrometry 
yields an even better positional coincidence of  
the 3--7\,keV peak with the West radio nucleus (Iwasawa et al.\ 2005).
The iron K$\alpha$ 
emission at 6.7\,keV with a large equivalent width (1.9\,keV), 
found by {\it XMM-Newton}, may also indicate a powerful AGN, hidden by 
high H$_2$ column densities (Iwasawa et al.\ 2005).

\noindent
2) {\it Evidence for high column densities that could hide an AGN:}
Numerous radio line interferometer maps show a high gas density toward
the West nucleus (e.g., Baan \& Haschick 1995 (BH95), Scoville et al.\
1997; Downes \& Solomon 1998 (DS98); Sakamoto et al.\ 1999; Mundell et
al.\ 2001).  High-resolution mid-IR maps from the Keck telescope show
a warm dust source in the West nucleus that is opaque at 25\,$\mu$m
(Soifer et al.\ 1999).  On a larger spatial scale than the West
nucleus, the global Arp~220 continuum spectrum measured by the ISO-LWS
indicates a dust $\tau > 1$ at 100\,$\mu$m, implying an H$_2$ column
density of $>2.7\times 10^{25}$\,cm$^{-2}$ (Fischer et al.\ 1997).
The OH and H$_2$O lines observed by ISO-LWS (Gonz\'alez-Alfonso et
al.\ 2004), and the strong deficiency in the PAH 7.7\,$\mu$m strength
vs.\ 850\,$\mu$m flux (Haas et al.\ 2001; Soifer et al.\ 2002; 
Spoon et al.\ 2004) also
imply an extinction large enough to hide the hard X-ray emission from
an AGN accretion disk.

\noindent
3) {\it Plausible cm-VLBI candidates for an AGN} in the West nucleus
are the flat-spectrum sources sources W10, W17, and W42, and the
rapidly-varying source W33 (Parra et al.\ 2007; Lonsdale et al.\ 2006).
VLBA monitoring of the radio supernovae within the Arp~220 nuclei
suggest that the supernova rate and starburst efficiency should be revised
downward, which would lower the starburst contribution
to the total luminosity (Rovilos et al.\ 2005; but see Lonsdale et al.\ 2006
and Parra et al.\ 2007).

\noindent
4) {\it Kinematic data:} The CO has a very high velocity spread
in the West nucleus.  One of the OH masers has a high velocity gradient
possibly marking the site of an AGN (Rovilos et al.\ 2003).
Broad ammonia absorption, with a velocity
spread of 700\,\kms , suggests molecular material in a small, rapidly
rotating disk surrounding a black hole (Takano et al.\ 2005).  While
the large-scale (15 to 28 kpc) ionized gas is dominated by tidal
motions rather than galactic winds (Colina et al.\ 2004), the complex
kinematics of the ionized gas in the central 2 kpc is influenced by
outflows from the dust-enshrouded nucleus --- see the H$\alpha$ and
[N~II] results (Arribas et al.\ 2001), and the {\it
Chandra} data (McDowell et al.\ 2003).

\begin{figure} \centering
\includegraphics[angle=-90,width=8.5cm]{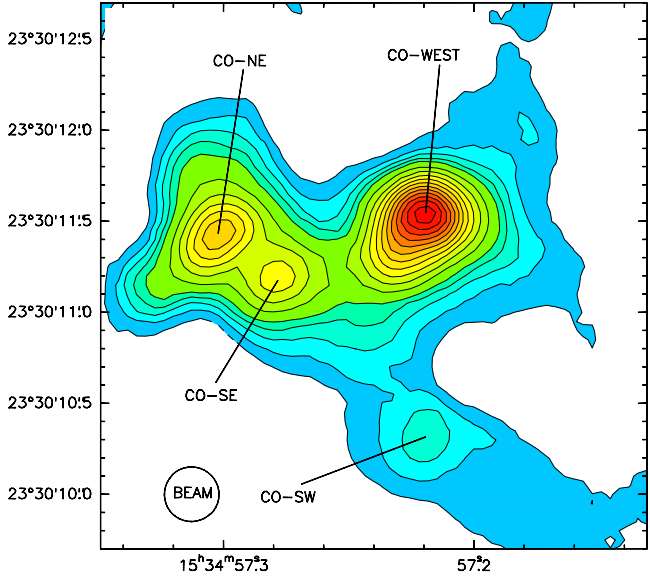}
\caption[CO21 Map]
{The central $3''$ of Arp~220 in CO(2--1) integrated over 770\,\kms ,
with the 1.3\,mm continuum subtracted.  The beam ({\it lower left}) is
$0.30''$ with $T_b/S = 266$\,K/Jy.  Contours are 2 to 10 by 2, then 14
to 54 by 4 (in Jy\,\kms ).  The CO-West peak is 56.5\,Jy\,\kms ; CO-NE
is 33.4\,Jy\,\kms .
}
\label{1mmlsbmap} 
\end{figure}

\begin{figure} \centering
\includegraphics[angle=-90,width=8.5cm]{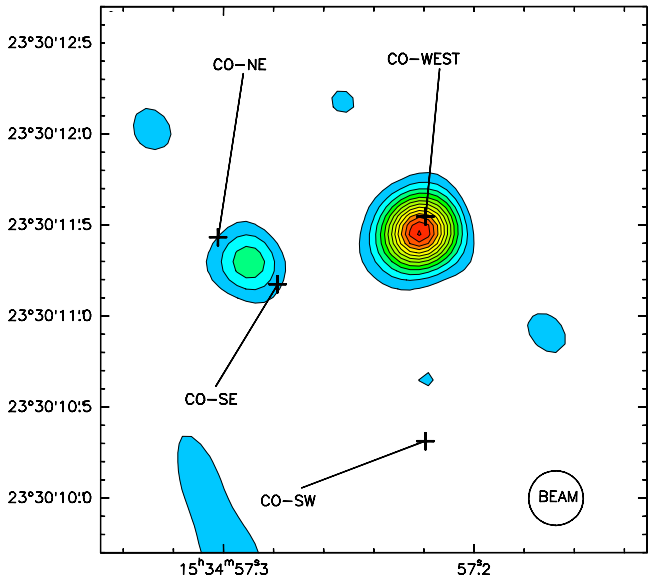}
\caption[1.3mmcont map]
{Continuum map at 1.3mm (229.4\,GHz). Contour steps are
6\,mJy\,beam$^{-1}$.  The Arp~220-West peak is 79\,mJy\,beam$^{-1}$,
and the East peak is 23\,mJy\,beam$^{-1}$.  Note that the continuum
peaks do not coincide with the CO(2--1) peaks, which are marked with
crosses.  The beam is 0.30$''$ (lower right).}
\label{1mmcont} 
\end{figure}

\section{New long-baseline observations}
To further investigate Arp 220's power source, we re-observed the
millimeter continuum and the CO(2--1) and (1--0) lines with the IRAM
Plateau de Bure interferometer with its new
long baselines to 760\,m, which enabled us to restore the data with
uniformly-weighted synthesized circular clean beams of 0.30$''$ at
1.3\,mm and 0.60$''$ at 2.6\,mm.  We calibrated 
amplitudes and phases with 3C273, 0923+392, 3C345, and 
1502+106.  Receiver temperatures were 45 to
65\,K at both wavelengths. 

This Letter reports on the 1.3\,mm results on the West nucleus.
We observed the 1.3\,mm continuum in the receivers' upper sideband
simultaneously with CO(2--1) in the lower sideband.  The spectral
correlators covered 770\,\kms\ at 1.3\,mm, with a channel spacing of
1.66\,\kms .  Our velocity scales are relative to 226.422\,GHz, which
is the rest frequency of CO(2--1) divided by 1+$z_{\rm lsr}$, where we
took $cz_{\rm lsr}$ = 5450\,\kms\ as the cosmological redshift.
Toward Arp~220, $V_{\rm lsr} = V_{\rm hel} + 16.6$\,\kms , so zero
velocity offset on our spectra is 5450\,\kms\ (LSR) and 5433\,\kms\
(heliocentric).

\begin{figure} \centering
\includegraphics[angle=-90,width=8.5cm]{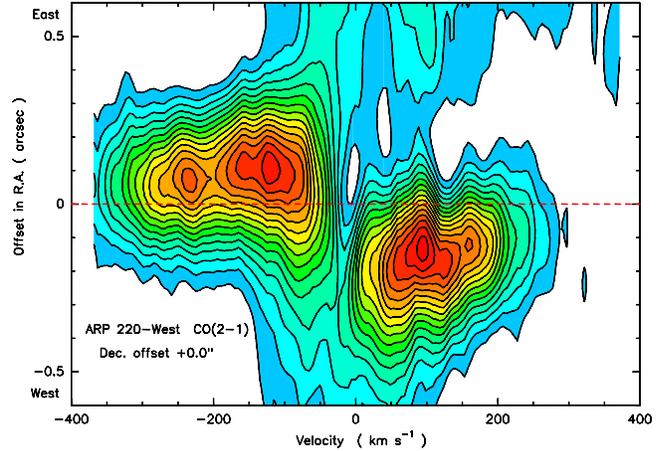}
\caption[CO21 pv diagram]
{East-west CO(2--1) position-velocity cut, through the West nucleus.
CO(2--1) contours are in steps of 10\,mJy\,beam$^{-1}$, with the
1.3\,mm continuum subtracted. The CO ring around the West nucleus
covers velocities from $-$370 to +300\,\kms , and is cut by absorption
at $-60$ to +40\,\kms .  Velocity offsets are relative to 226.422\,GHz
($cz_{\rm lsr}$ = 5450\,\kms ).  R.A.\ offsets are relative to the
West continuum peak, indicated by the horizontal line (position in
Table~1).
}
\label{co21-pv} 
\end{figure}

\begin{table*}
\caption{Positions, Sizes, and fluxes of the Arp~220-West nuclear disks.}
\begin{tabular}{lll cccc ccc}
\hline
 &R.A. 	  &Dec.	 &Major  &Minor	&P.A.  &Observed	
&\multicolumn{3}{c}{Model Components}\\
 &15$^{\rm h}$34$^{\rm m}$
	  &23$^\circ 30'$  &axis &axis &  &flux density &Dust &H II &non-th.
\\ 
{\bf Data} &(J2000)  &(J2000) &(arcsec) &(arcsec) &(deg.) &(mJy) 
&(mJy) &(mJy) &(mJy)
\\	     
\hline
\multicolumn{5}{l}{ {\bf Arp~220-West Continuum:} }\\
West 1.3\,mm   
        &57$^{\rm s}.2226$   
	&$11.46''$ 
	&$0.19''\pm 0.01''$
	&$0.13''\pm 0.02''$
	&$-37^\circ\pm 6^\circ$
	&$106\pm 2$
	&94
        &10
        &2.7
\\
West 2.6\,mm 
        &57$^{\rm s}.2221$   
	&$11.48''$ 
	&$0.14''\pm 0.08''$
	&---
	&---
	&$25\pm 1$
	&8.5
        &11
        &4.7
\\
\multicolumn{5}{l}{ {\bf West-CO(2--1)  $-$380 to $-$70\,\kms:} }
        &&(Jy\,\kms )
\\
	&57$^{\rm s}.2265$  
	&$11.48''$
	&$0.37''\pm 0.1''$
	&$0.39''\pm 0.1''$
	&$-32^\circ\pm 5^\circ$
	&74.2
        &---
        &---
        &---
\\
\multicolumn{5}{l}{ {\bf West-CO(2--1) +40 to +380\,\kms :} }
\\
	&57$^{\rm s}.2157$  
	&$11.51''$
	&$0.32''\pm 0.1''$
	&$0.22''\pm 0.1''$
	&$-80^\circ\pm 5^\circ$
	&49.3
        &---
        &---
        &---\\

\hline  
\multicolumn{8}{l}{Estimated position errors  
 are $\pm 0.004^{\rm s}$ in R.A.\ and $\pm 0.05''$ in Dec.}
\\  
\end{tabular}
\end{table*}

The new CO(2--1) map (Fig.\,1) shows four peaks, plus an
extension of the CO-West gas to the southeast (``source C'' on the
maps by BH95).  The 1.3\,mm continuum map (Fig.\,2) yields a
source position (Table~1) that agrees within the errors with
our earlier IRAM data (DS98).  But an important new result is the small
measured size of the West continuum, which is 0.19$''\times 0.13''$.
This means that the continuum, which at 1.3\,mm is mainly dust
emission, {\it does not trace the same matter as the CO(2--1)}.  The
dust is more compact than the CO emission, which appears to be in
a larger ring or disk around the compact dust core. This is clearly 
shown in the east-west position-velocity cut through the West nucleus
(Fig.\,3), which covers the full velocity range of 700\,\kms\ in
the CO, but with a dramatic change from negative to positive
velocities over the central 0.2$''$, where the CO contours are cut by a deep
absorption trough.  The CO absorption is even more spectacular in the
individual spectra in steps of 0.1$''$ across the West nucleus 
(Fig.\,4).  The main absorption appears to be 100\,\kms\ wide and
centered on $-10$\,\kms , with some of the spectra showing a second
absorption feature at +130\,\kms .  Modeling suggests this is partly
absorption of the hotter continuum, and CO self-absorption, implying
that the CO-West disk itself has a temperature gradient increasing
inwards.  This is the first time that this Arp~220-West absorption has been
seen in CO, and is mainly due to our improved spatial resolution.  In
previous larger-beam observations, the CO absorption has probably been
masked by CO emission in the beam.  The CO absorption 
probably also explains why the   
West continuum peak is at a slightly different position than the CO 
(Fig.~2 and Table~1).

\begin{figure} \centering
\includegraphics[angle=-90,width=8.5cm]{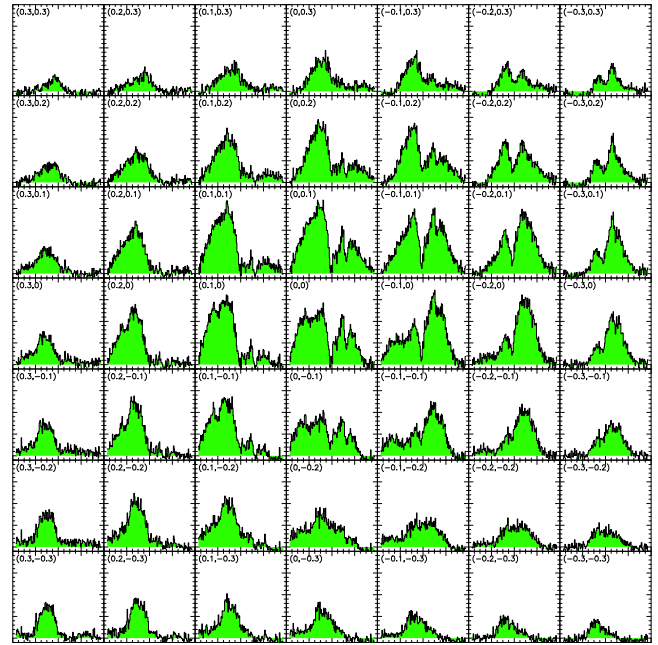}    
\caption[CO(2-1) Spectra]{
Evidence for foreground obscuration: CO(2--1) spectra across
Arp~220-West, with the 1.3\,mm continuum subtracted.  Each box is a
step of 0.1$''$, north is up, east is left, and (R.A.,Dec). offsets
are in the upper left of each box.  The grid center at (0,0) is the
1.3\,mm West continuum peak (Table~1).  Velocities run from $-$400 to
+400\,\kms , the spectral resolution is 5\,\kms , and zero velocity
(the center of the spectra) is at 226.422\,GHz ($cz_{\rm lsr} =$
5450\,\kms ).  Intensities run from 0 to 200\,mJy\,beam$^{-1}$.  The
beam is $0.30''$ with $T_b/S = 266$\,K/Jy, which means the
directly-observed, beam-smoothed, peak CO brightness temperatures are
40\,K. Note the deep absorption near the line center in some of the
spectra, which indicates foreground, cooler gas.
}
\label{a220W-CO21-spectra} 
\end{figure}

\section{West dust disk: compact, hot and opaque.}
The West continuum peak is 79\,mJy in our 0.3$''$ beam, and its
spatially integrated flux is 106\,mJy, as in our earlier IRAM result
(DS98).  From lower-frequency data (Sopp \& Alexander 1991;
Anantharamaiah et al.\ 2000; Rodriguez-Rico et al.\ 2005), we estimate
that at 1.3\,mm, the West nucleus has an extended synchrotron flux of
2.7\,mJy, and a free-free continuum of 10\,mJy (Table~1;
Fig.~5).  Extrapolating the 3.6\,cm flux of 400\,$\mu$Jy of the
variable VLBI source W33 (Parra et al.\ 2007) with an assumed spectral
index of +0.3, as in Sgr~A$^*$, we estimate that any synchrotron
self-Compton contribution is $< 2$\,mJy at 1.3\,mm.  Most of the
1.3\,mm flux must therefore be dust emission, at a level of $\sim
94$\,mJy.  Hence, the directly-observed, beam-smoothed, dust
brightness temperature is 18\,K, a remarkably high value at 1.3\,mm.
Our Gaussian fits to the West continuum in the $(u,v)$-plane yield a size
of $0.19''\times 0.13''$, so the deconvolved dust brightness
temperature is a spectacular 90\,K at 1.3\,mm.  The Arp~220 west
nucleus is thus a very unusual dust source, not at all like the cooler
dust sources detected at millimeter wavelengths in starburst
galaxies.

The small source size at 1.3\,mm and the deconvolved dust
brightness temperature already imply that the 1.3\,mm dust opacity is
unusually high.  In the simplest estimate, $\tau = - \ln (1 -
T_b/T_d)$.  If we assume that the intrinsic dust temperature is in the
range (90$<T_d\leq 180$\,K), so as not to exceed the Arp~220
IRAS FIR luminosity, then $\tau$(1.3\,mm) $\geq 0.7$.  This lower limit is
already quite a ``high'' opacity at 1.3\,mm.  Such a high-brightness
dust source at 1.3\,mm will certainly be opaque in the far-IR, so its
intrinsic far-IR SED is simply a Planck curve.  The spectrum deviates
from a blackbody curve on the Wien side however, because it is
attenuated by foreground dust.  We observe deep CO(2--1) absorption
due to foreground material, so this foreground attentuation must
exist.  The foreground dust is not as dense as the West-nucleus dust
itself; it is optically thin at millimeter wavelengths, but optically
thick in the mid-IR.  Hence to fit the observed fluxes from the
submillimeter through the mid-IR with a blackbody curve, we must first
de-redden the observed mid-IR fluxes measured at the Keck telescope by
Soifer et al. (1999). In a first iteration, we tried to match the 
luminosity of the West nucleus derived independently by Soifer et al.\
and Haas et al. (2001), and started with the foreground
dust opacity ($\tau$ = 1.2 to 1.5) at 25\,$\mu$m that Soifer et al.\
estimated by {\it assuming} the source size and dust temperature 
to be in the range from  (0.39$''$, 102\,K) to (0.25$''$, 128\,K).  
The source size that we
measure, however, is even smaller, $0.19''\times 0.13''$,
implying a higher intrinsic
dust temperature to reach the same luminosity, 
and hence a slightly higher foreground opacity
to fit the mid-IR data points.  Our current best compromise is shown
in Fig.~5.  This solution is for an intrinsic  dust temperature of
170\,K, a foreground dust opacity of $\tau$ = 1.7 at 25\,$\mu$m, with
a $\lambda^{-1}$ opacity dependence at shorter wavelengths, and a
$\lambda^{-2}$ dependence at longer wavelengths.

In this solution, the Arp~220-West dust that we observe at millimeter
wavelengths (not the foreground dust) has an optical depth of unity at
1.1\,mm.  For the observed source size, and an intrinsic dust
temperature of 170\,K, we obtain a total IR luminosity of $9\times
10^{11}$\,\Lsun\ for the West nucleus.  The dust flux implies a a gas
mass (H$_2$+He) of $1\times 10^9$\,\Msun , and a mean H$_2$ density of
$9\times 10^4$\,cm$^{-3}$, or 5000\,\Msun\,pc$^{-3}$.  The mass could
be lower if the dust grains are unusually large, and/or the abundances
are super-solar.  Such an enclosed mass yields, at radius 30\,pc, a
rotation velocity $(GM/R)^{0.5}$ of 370\,\kms , and this is about what
we observe in CO.  For this West dust source (not the foreground
dust), if the optical depth scales as $\lambda^{-2}$ to the mid-IR,
then we expect $\tau = 2000$ at 25\,$\mu$m.  If $\tau\sim\lambda^{-1}$
from the mid-IR to the visible, then we expect $\tau \approx 10^5$ at
5000\,\AA .

\begin{figure} \centering
\includegraphics[angle=-90,width=8.5cm]{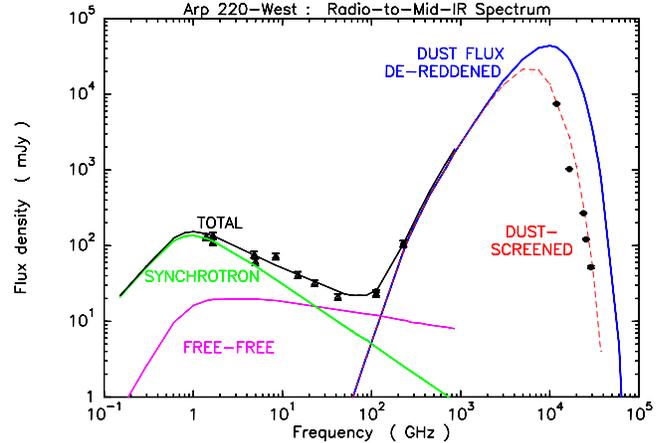}
\caption[Continuum SED]{
Our model of the radio-to-mid-IR continuum spectrum of
Arp~220-West. The spectrum is dominated by an opaque, 170\,K blackbody
dust component, with a luminosity of 9$\times 10^{11}$\,\Lsun , that
only becomes optically thin below 1.1\,mm.  The dust blackbody is
absorbed by a foreground screen (the West-CO ring and possibly part 
of the Eastern disk) 
that has a foreground opacity of 1.7 at 25\,$\mu$m.  The
effect of the foreground screen is to attenuate the intrinsic mid-IR
fluxes, thus shifting the apparent peak of the dust blackbody to
longer wavelengths, giving the illusion of a lower temperature (dashed
curve).  Besides the millimeter-to-mid-IR dust component, there is a a
free-free component that becomes optically thick below 20\,cm, and
extended synchrotron emission mixed with the ionized gas, that turns
over below 20\,cm due to free-free absorption.  The data points at 113
and 229 GHz are from this paper.  The other radio fluxes are from the
references cited in the text, and the mid-IR points are the 10 to
25$\mu$m fluxes measured by Soifer et al.\ (1999).
}
\label{a220W-continuum-SED} 
\end{figure}

What about the foreground dust? The observed CO absorption shows that the
West dust core is also obscured by its surrounding, rapidly rotating
West molecular disk,  and 
possibly by some of the off-plane, larger-scale, Eastern disk
material that may be in front of the West nucleus (see sketch by
Mundell et al.\ 2001).  This Eastern disk gas, that we modeled as a
warped disk with a quasi polar-ring structure (Eckart \& Downes 2001),
is resolved, and therefore mostly absent 
in the long-baseline data presented here, but it may
contribute to the absorption features that cut into the CO(2--1) line
profiles from the West nucleus.  For the values in our dust model
shown in Fig.~5, the {\it observed} flux at 25\,$\mu$m is
7.5\,Jy (Soifer et al.\ 1999).  The {\it intrinsic}, de-reddened dust
flux at 25\,$\mu$m is 42\,Jy.  The difference, due the foreground dust
of the West molecular disk and possibly part of 
the Eastern disk, corresponds to a 
foreground opacity of 1.7 at 25\,$\mu$m, close to the value of 1.5
estimated in one of the models by Soifer et al.  It also implies a
foreground opacity of $\sim$20 at {\it K} band, and a foreground A$_V
>$ 100 in the visible, as previously inferred by Haas et al.\ (2001)
from the weakness of the PAH feature.  {\it This foreground extinction
is why the Arp~220 has a cooler SED than the ``warm'', AGN-powered
ULIRGs} like Mrk~231 and the nearby quasars. At visible wavelengths,
the AGN is hidden by 10$^4$ to 10$^5$\,mag of obscuration by the dust
torus, and this dust torus's radiation is itself attenuated by an
additional 100\,mag due to the compact, 0.5$''$ West CO disk {\it and}
possibly part of the larger-scale, warped Eastern CO disk.

\section{Conclusions}
New CO(2--1) and 1.3\,mm continuum data
provide more evidence for a black hole in the Arp 220-West
nucleus:

\noindent
1) The Arp 220-West dust continuum has a {\it 1.3\,mm brightness
 temperature of 90\,K.} This is much hotter than the dust detected at
 millimeter wavelengths from starburst galaxies. SED fitting implies a
 dust opacity of unity at 1.1\,mm, and a true dust temperature of
 170\,K, so the West disk strongly resembles that of the compact dust
 toroid around the quasar APM 08279+52 (Egami et al.\ 2000; Weiss et al.\
 2007).

\noindent
2) The size of the West dust source is $35 \times 20$\,pc.
 This size and the 1.3\,mm dust flux imply a gas density
$>5000$\,\Msun\,pc$^3$, about 10 times the stellar density 
in cores of giant ellipticals. Model SED fits that attempt to 
correct upward the mid-IR fluxes for attenuation by the 
foreground absorbing screen  are consistent with 
a bolometric luminosity of $9\times 10^{11}$\,\Lsun , that is, 
 75\% of the total IRAS luminosity of Arp~220. Because of the 
foreground screen, the true {\it bolometric} luminosity is unknown;
depending on geometry, 
it may be significantly 
higher than the IR luminosity derived from the IRAS data.
 
\noindent
3) Strong CO absorption is seen in front of the dust continuum source.
 The inner dust source is a hot region, and is not the same source as
 the surrounding, cooler CO.

\noindent
4) The West-CO torus centered on the compact 1.3\,mm dust source has a
steep velocity rise toward the nucleus, which is characteristic of a
massive black hole (Fig.~3).  The West molecular gas does not
follow a rotation curve rising with radius, typical of inner-galaxy
bulge regions.  The CO must be rotating in the gravitational potential
of a centrally-concentrated mass.  The CO velocities in the West
molecular torus extend to 370\,\kms\ at a radius of 30\,pc, which
argues for an enclosed mass (gas + stars + black hole) of $1\times
10^9$ \Msun .

\noindent
5) The data are consistent with the CO being in a cooler (50\,K) ring
 surrounding a much hotter (170\,K), dense dust source.  The hot dust
 source is an inner, probably self-gravitating disk of radius 35\,pc.
 The radio supernova candidates seen in VLBI maps extend over a
 slightly larger region, and some of them may be in the West CO torus
 rather than in the inner, opaque dust disk.

\noindent
6) The combined proton column densities from the 
foreground main Arp 220 CO
disk (10$^3$\,cm$^{-3}$ x 300\,pc), 
plus the dense CO-West torus (10$^4$\,cm$^{-3}$ x 90\,pc), 
plus the very dense, innermost dust disk (10$^5$\,cm$^{-3}$ x 30\,pc)
add up to $\sim 1.3\times 10^{25}$\,cm$^{-3}$.  This is sufficient to
hide all of the optical emission and most of the hard X-ray
emission from the supermassive black hole accretion disk, which must
be at the center of the 0.19$''$ continuum source seen at 1.3\,mm.

\noindent
7) {\bf Why it is a black hole:} The dust source seen in these
millimeter interferometer observations is small (0.19$''\times
0.13''$) and optically thick at 1.1\,mm.  Its blackbody luminosity is
nearly 10$^{12}$\,\Lsun .  The area of the dust disk on the sky is
$2\times 10^3$\,pc$^2$, so the emission surface brightness is $\sim
5\times 10^{14}$\,\Lsun\,kpc$^{-2}$.  This puts impossible
requirements on a starburst: the equivalent of 10 million O~stars
packed into the $r=35$\,pc dust disk, with $\sim 400$ O~stars in each
cubic parsec, or 30 times the luminosity of M82 from a thousand times
smaller volume.  No such super-starburst has ever been observed.  This
means the true source of the Arp~220-West luminosity cannot be a
starburst. It can only be a black hole accretion disk.

\begin{acknowledgements}
We thank the Plateau de Bure Interferometer operators for their help
with the observing, C.\ Thum, IRAM-Grenoble, 
for useful discussions, J.~Conway, Onsala Space Observatory, for comments, 
and the referee
for very helpful suggestions on improving the paper.  IRAM is
supported by INSU/CNRS (France), MPG (Germany) and IGN (Spain).
\end{acknowledgements}

\end{document}